\documentclass[twocolumn]{revtex4}
\usepackage{amsmath,amssymb,graphicx,bbm}
\usepackage{times}

\newcommand{\tmop}[1]{\ensuremath{\operatorname{#1}}}
\newcommand{\um}{-}

\begin{document}

\title{Polarization Suppression and Nonmonotonic Local Two-Body Correlations in the Two-Component
Bose Gas in One Dimension} 

\author{Jean-S\'ebastien Caux${}^1$, Antoine Klauser${}^{1,2}$ and Jeroen van den Brink${}^2$}
\affiliation{${}^1$Institute for Theoretical Physics, Universiteit van  
Amsterdam, 1018 XE Amsterdam, The Netherlands}
\affiliation{${}^2$Instituut-Lorentz, Universiteit Leiden, P. O. Box 9506, 2300 RA Leiden, The Netherlands}

\date{\today}

\begin{abstract}
We study the interplay of quantum statistics, strong interactions and finite temperatures in 
the two-component (spinor) Bose gas with repulsive delta-function interactions in one dimension.  
Using the Thermodynamic Bethe Ansatz, we obtain the
equation of state, population densities and local density correlation numerically as a function of 
all physical parameters (interaction, temperature and chemical potentials), quantifying the full crossover 
between low-temperature ferromagnetic and high-temperature
unpolarized regimes.  In contrast to the single-component, Lieb-Liniger gas,
nonmonotonic behaviour of the local density correlation as a function of temperature is observed.
\end{abstract}

\maketitle

The experimental realization of interacting quantum systems using cold atoms
has reignited interest in many-body physics of strongly-interacting quantum systems in and out of equilibrium
\cite{blochRoMP}.  Effectively one-dimensional realizations of bosonic $^{87}\mbox{Rb}$ quantum gases with tunable local interaction 
strength \cite{MoritzPRL91,LaburtheTolraPRL92,ParedesNATURE429,KinoshitaSCIENCE305,PolletPRL93} realize
the single-component Lieb-Liniger model \cite{LiebPR130,LiebPR130b}, for which 
the crossover from weakly- to strongly-interacting physics is experimentally accessible and well understood
from first principles.  Observed finite temperature thermodynamics \cite{AmerongenPRL100} even fit
predictions from the Thermodynamic Bethe Ansatz (TBA) \cite{yang&yang}. 
For a single bosonic species in one dimension, statistics and interactions are intimately related: the limit of 
infinitely strong interactions (impenetrable bosons)
causes a crossover from bosonic to effectively fermionic behaviour \cite{TonksPR50,GirardeauJMP1}
for density-dependent quantities.  Density profiles and fluctuations
accessible from exact thermodynamics allow to discriminate between 
these fermionized and quasicondensate regimes \cite{PetrovPRL85,GangardtPRL,KheruntsyanPRL,KheruntsyanPRA}. 
Multicomponent (spinor) systems however provide a much larger number of different regimes than their single-component
counterparts, and realize situations where important interaction and quantum statistics effects coexist
and compete.  Their thermodynamics has not been extensively studied using exact methods;  in this work,
we wish to highlight some unexpected features inherent to a system in this class.

The experimentally realizable \cite{widera:140401} case of two-component bosons in 1D with symmetric interactions, which we will focus on here, 
contrasts with the Lieb-Liniger case in many ways.  The ground state is (pseudo-spin) polarized \cite{SutherlandBOOK,Yang&Li} (ferromagnetic), 
as expected from a general theorem valid when spin-dependent 
forces are absent \cite{EisenbergPRL89}, and thus coincides with the Lieb-Liniger ground state.  
On the other hand, excitations carry many additional branches, starting from the simplest spin-wave-like one.  
These excitations are difficult to describe in general, even in the strongly-interacting limit 
(there, no effective fermionization can be used, since the two pseudospin components remain strongly coupled),
where spin-charge separation occurs \cite{FuchsPRL,LiEPL61,BatchelorJoSM,ZvonarevPRL99,Kleine}. 
The thermodynamic properties are drastically different from those of the one-component Lieb-Liniger gas \cite{GuIJMPB16} and at large 
coupling and low temperature correspond to those in an isotropic $XXX$ ferromagnetic chain \cite{GuanPRA76}. 
Temperature suppresses the entropically disfavoured polarized state, and opens up the possibility of 
balancing entropy and quantum statistics gains (from the wavefunction symmetrization requirements) 
with interaction and kinetic energy costs in the free energy.
Using a method based on the integrability of the system, we find that this thermally-driven interplay
leads to a correlated state with many interesting features, the most remarkable being a nonmonotonic dependence of the local density
fluctuations of the system with respect to temperature or relative chemical potential.  

For definiteness, we consider a system of $N$ particles on a ring of length $L$, 
subjected to the Hamiltonian 
\begin{equation}
  \mathcal{} \mathcal{H}_N =^{} - \frac{\hbar^2}{2m} \sum^N_{i = 1} \frac{\partial^2}{\partial
  x^2_{i^{}}} + g_{1D} \sum_{1 \leq i < j \leqslant N} \delta (x_i - x_j).
\label{eq:H_N}
\end{equation}
The effective one-dimensional coupling parameter $g_{1D}$ is related to the effective 1D scattering length $a_{1D}$
\cite{OlshaniiPRL81} via the relation $g_{1D} = \hbar^2 a_{1D}/2m$, and to the effective interaction
parameter $\gamma = c/n$ (where $n = N/L$ is the total linear density) via $c = g_{1 D} m/\hbar^2$. 
We set $\hbar = 2m = 1$ to simplify the notations.  
%Note that this involves fine-tuning one parameter:  in more generality, we could have different intra- and inter-species scattering
%lengths.  To preserve integrability however, these must all be equal.
Yang and Sutherland  \cite{YangPRL19,SutherlandPRL20}
showed that the repulsive delta-function interaction problem admits an exact solution irrespective 
of the symmetry imposed on the wavefunction, meaning that the wavefunctions of (\ref{eq:H_N})
are of Bethe Ansatz form whether the particles are distinguishable, or mixtures of various bosonic and fermionic species
\cite{GaudinBOOK}.
The ground state and excitations of multicomponent fermionic system were studied, both for repulsive and attractive interactions, by Schlottmann
\cite{SchlottmannJPC5,SchlottmannJPC6},
but these results cannot be translated to the bosonic case we are interested in.

Specifically, specializing to $N$ atoms of which $M$ have (in the adopted cataloguing) spin down, the Bethe Ansatz
provides eigenfunctions fully characterized by sets of rapidities (quasi-momenta) $k_j$, $j = 1, ..., N$ and pseudospin rapidities
$\lambda_{\alpha}$, $\alpha = 1, ..., M$, provided these obey the $N + M$ coupled Bethe equations \cite{YangPRL19,SutherlandPRL20}
\begin{eqnarray}
&&  e^{\tmop{ik}_j L}  =  - \prod_{l = 1}^N \frac{k_j - k_l + \tmop{ic}}{k_j -
  k_l - \tmop{ic}} \prod^M_{\alpha = 1} \frac{k_j - \lambda_a -
  \frac{\tmop{ic}}{2}}{k_j \um \lambda_a + \frac{\tmop{ic}}{2}}, \nonumber \\
%&& \hspace{5cm} j = 1,  \ldots, N \nonumber\\
&&  \prod^N_{l = 1} \frac{\lambda_{\alpha} - k_l -
  \frac{\tmop{ic}}{2}}{\lambda_{\alpha} - k_l + \frac{\tmop{ic}}{2}} = -
  \prod^M_{\beta = 1} \frac{\lambda_{\alpha} - \lambda_{\beta} -
  \tmop{ic}}{\lambda_{\alpha} - \lambda_{\beta} + \tmop{ic}}, 
%  \nonumber \\
%&& \hspace{5cm} \alpha = 1, \ldots, M 
\label{Bethe_eq}
\end{eqnarray}
for $j = 1, ..., N$ and $\alpha = 1, ..., M$.  For a generic eigenstate, the solution to the Bethe equations
is rather involved.  In general, the $k_j$ rapidities live on the real axis; the $\lambda_{\alpha}$ are on the other
hand generically complex, but arranged into regular string patterns, each type of string representing
a distinct type of quasiparticle.  The spectrum of the theory thus contains infinitely many branches, with an
increasing effective mass.  In the continuum limit $N \rightarrow \infty, L \rightarrow \infty, N/L$ fixed,
the TBA allows to exploit the condition of thermal equilibrium to 
obtain the Gibbs free energy of the system as a function of the temperature $T$ and of the
total $\mu$ ($=\frac{\mu_1+\mu_2}{2}$ with $\mu_i$ the chemical potential specific to the $i$th component) and relative $\Omega$ ($=\frac{\mu_1-\mu_2}{2}$) chemical potentials.  As detailed for example in \cite{TakahashiBOOK}
the Gibbs free energy density is given by 
\begin{equation}
g = -T \int_{-\infty}^{\infty} \frac{dk}{2\pi} \ln \left[ 1 + e^{-\epsilon(\lambda)/T} \right]
\label{eq:Gibbs}
\end{equation}
where $\epsilon (\lambda)$ is dressed energy, which is self-consistently coupled
to the length-$n$ string dressed energy $\epsilon_n (\lambda)$, $n = 1, 2, ...$ via the system
\begin{eqnarray}
  \varepsilon (\lambda) \!\!&=&\!\! \lambda^2 - \mu - \Omega - T a_2 \!\ast\! \ln
  \left[ 1 + e^{-\varepsilon(\lambda)/T} \right] \nonumber \\
  && - T \sum_{n = 1}^{\infty} a_n \!\ast\! \ln \left[ 1 + e^{- \varepsilon_n(\lambda)/T} \right] \\
  \frac{\varepsilon_1 (\lambda)}{T} \!\!&=&\!\! f \!\ast\! \ln \left[ 1 + e^{-\varepsilon(\lambda)/T} \right] + f \!\ast\! \ln \left[ 1 + e^{\varepsilon_{2}(\lambda)/T} \right], \\
  \frac{\varepsilon_n (\lambda)}{T} \!\!&=&\!\! f \!\ast\! \ln \left[ 1 + e^{\varepsilon_{n + 1}(\lambda)/T} \right] + f \!\ast\! \ln \left[ 1 +
  e^{\varepsilon_{n - 1}(\lambda)/T} \right],
  \nonumber \\ &&\hspace{5cm} (n > 1),
\label{system}
\end{eqnarray}
with the standard convolution notation $g \ast h (\lambda) \equiv \int_{-\infty}^{\infty} d\lambda' g (\lambda - \lambda')
h (\lambda')$, and the kernels $a_n (\lambda) = \frac{1}{\pi} \frac{n c / 2}{(n c / 2)^2 + \lambda^2}$ and
$f (\lambda)  = \frac{1/2c}{\cosh ( \pi \lambda/c)}$.  This set of coupled equations is complemented
with the asymptotic condition $\lim_{n \rightarrow \infty} \frac{\varepsilon_n (\lambda)}{n} = 2 \Omega$.
%for high-level functions.  % and asymptotic solutions in rapidity ${k \rightarrow \infty}$. 
In view of its nonlinear nature, we solve the infinite system of coupled integral equations numerically.
In order to validate our results, we have in fact implemented two independent algorithms, one based on fast Fourier transforms
and the other based on adaptive-lattice integration specifically tailored to the computation of the free energy or of its various derivatives.
While extremely challenging, it remains possible to obtain good accuracy throughout parameter space, except
in the extreme limit of vanishing relative chemical potential for  $\Omega<<T, \mu, c$ or the low 
nonzero temperature limit $0<T<<\Omega, \mu, c$ \footnote{Because of the covariance of the 
theory under overall rescaling of $T, \mu, \Omega$ and $c$, we can always perform calculations using $c = 1$.}.
Leaving the details to a future publication, we here simply concentrate on the results obtained.

\begin{figure}[ht]
\includegraphics[width=8.5cm]{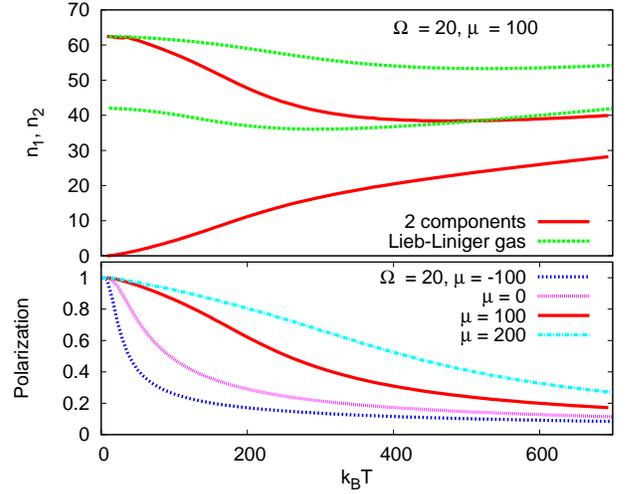}
\caption{Population densities (top) and polarization (bottom) of the spinor Bose gas as a function of temperature,
for fixed chemical potentials, and contrasted to separate Lieb-Liniger gases at
corresponding chemical potentials.  The Lieb-Liniger result for the majority 
chemical potential is recovered only at zero temperature, when ferromagnetism 
causes complete polarization.}
\label{fig:T}
\end{figure}
\begin{figure}
%\includegraphics[width=8.5cm]{2CBG_Om_mu_Pol.pdf}
%\includegraphics[width=8.5cm]{2CBG_Om_T_Pol.pdf}
%%\caption{Population densities as a function of the relative chemical potential
%%$\Omega$, at fixed total chemical potential $\mu$ and for different temperatures,
%%showing quickly increasing polarization in a large temperature window.}
%%\label{fig:Om_T}
%\caption{Population densities as a function of the relative chemical potential
%$\Omega$,firstly at fixed temperature and for different values of the total chemical
%potential $\mu$ and secondly at fixed total chemical potential $\mu$ and for different temperatures}
\includegraphics[width=8.5cm]{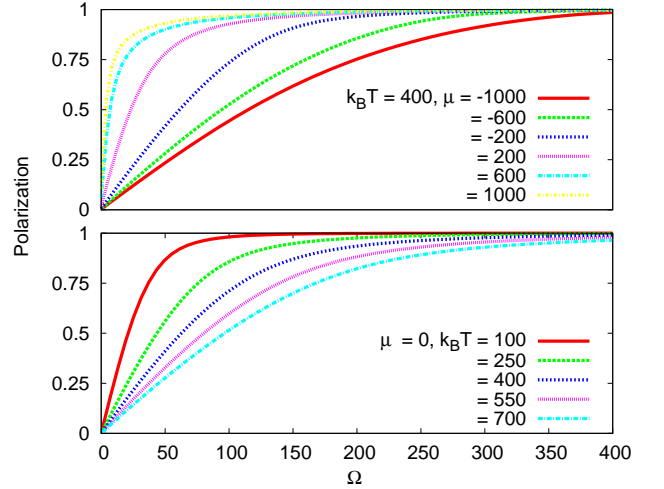}
\caption{Polarization as a function of the relative chemical potential
$\Omega$, (top) at fixed temperature and for different values of the total chemical
potential $\mu$ and (bottom) at fixed total chemical potential $\mu$ and for different temperatures.}
\label{fig:Om}
\end{figure}
%\begin{figure}[ht]
%\includegraphics[width=8.5cm]{2CBG_mu_T_Pol.pdf}
%\caption{Polarization as a function of the total chemical potential
%$\mu$, for fixed relative chemical potential $\Omega$ and for different
%temperatures.  Lowering the temperature increases the polarization at any
%$\mu$, leading back to Lieb-Liniger behaviour.}
%\label{fig:mu_T}
%\end{figure}
%\begin{figure}[ht]
%\includegraphics[width=8.5cm]{2CBG_mu_Om_Pol.pdf}
%\caption{Polarization of the spinor gas as a function of the total
%chemical potential $\mu$, for fixed temperature and for different values of
%the relative chemical potential $\Omega$.  The top curves give the densities
%of the majority component, the lower ones are for the minority one.}
%\label{fig:mu_Om}
%\end{figure}
\begin{figure}[ht]
\includegraphics[width=8.5cm]{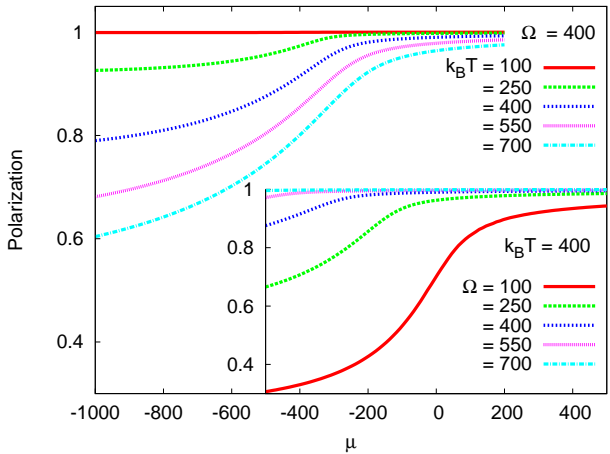}
\caption{Polarization as a function of the total chemical potential
$\mu$, for fixed relative chemical potential $\Omega$ and for different
temperatures (main) and (inset) for fixed temperature and for different values of
the relative chemical potential $\Omega$.  Lowering the temperature or increasing $\Omega$ 
increases the polarization at any $\mu$, leading back to Lieb-Liniger behaviour.}
\label{fig:mu_Pol}
\end{figure}

We focus on the polarization, defined as $(n_1-n_2)/(n_1+n_2)$ where $n_i=N_i/L$ the linear density of the $i$th 
boson component, and on the local density-density correlator
$g^{(2)}=\frac{\sum_{i,j}\langle \Psi^\dagger_i\Psi^\dagger_j \Psi_j\Psi_i \rangle}{\sum_{i}\langle \Psi^\dagger_i \Psi_i\rangle^2}$. 
The linear densities themselves, together with other equilibrium quantities such as the entropy, will be discussed
more extensively in a future publication.
Fig. \ref{fig:T} contrasts two single-component gases living in two different traps (Lieb-Liniger gases) against cohabitation in the same trap,
for a generic choice of chemical potentials.
The polarization of the ground state is clearly visible and gets suppressed with temperature at a rate depending on $\mu$, 
which itself sets the effective coupling $\gamma$.  In the limit of $\mu \ll 0$, $\gamma$ becomes bigger than $1/k_BT$, 
and the gas becomes paramagnetic.  For $\mu \gg 0$ ( $\gamma \ll 1$), the gas shows ferromagnetic behavior. 
The effect of the relative chemical potential on the polarization is shown in Fig. \ref{fig:Om}. 
Due to the finite temperature, the limit $\Omega \rightarrow 0$ is always unpolarized.
% and similarly to a magnetic system, the temperature reduce the alignment made by the relative chemical potential. 
Fig. \ref{fig:mu_Pol} shows data for a wide range of the total chemical potential. 
For $\mu$ small enough ($\gamma \gg 1$, paramagnetic), the value of the polarization depends only on $\Omega$ and $T$.
In contrast, for $\mu \gg 0$, the system behaves as a quasi-condensate ($\gamma \ll 1$, ferromagnetic).

\begin{figure}
\includegraphics[width=8.5cm]{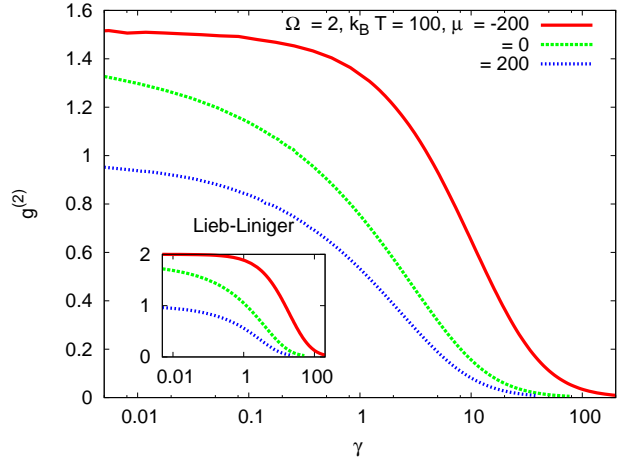}
\caption{Local pair correlation $g^{(2)}$ of the spinor Bose gas 
as a function of the effective coupling
$\gamma$, at fixed temperature and for three different values of the total chemical 
potential $\mu$.  The relative chemical potential $\Omega$ is set to a low value.  
Inset:  the same curves for the Lieb-Liniger gas.  The asymptotes $\gamma \rightarrow 0$
differ, but the general shape of the curve is very similar.}
\label{fig:C_c3}
\end{figure}

The observable $g^{(2)}$, which can be obtained from the interaction parameter derivative of (\ref{eq:Gibbs}),  
quantifies the fluctuations of density at a local point of space and time.
Fig. \ref{fig:C_c3} presents data for this quantity as a function of the effective interaction.  
In the limit $\gamma \rightarrow 0$ and in the decoherent regime (where the reduced temperature 
$\tau \equiv \frac{T}{(n_1 + n_2)^2} \gg 1$), the value saturates 
between $2$ (for $\Omega \rightarrow \infty$) and $1 + \frac{1}{N_c}$ (for $\Omega \rightarrow 0$, where $N_c = 2$ is the
number of components), generalizing the Lieb-Liniger result \cite{KheruntsyanPRL}. 
Our data fit well with this prediction, as seen in Fig. \ref{fig:C_c3}, where for $\gamma \sim 10^{-2}$, $g^{(2)}$ 
approaches $1.5$ for $\Omega$ and $\mu$ small.  
For bigger $\mu$, the data does not reach this value since the reduced temperature is too low and decoherent regime is thus not yet reached.
On the other hand, when the gas becomes strongly interacting, $g^{(2)}$ vanishes as expected.

\begin{figure}
\includegraphics[width=8.5cm]{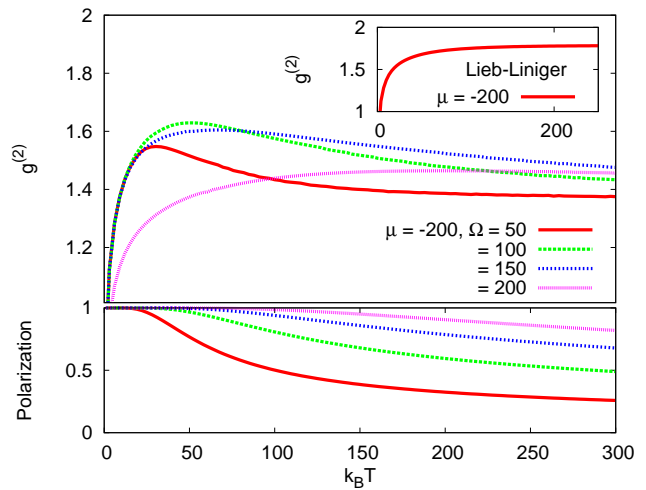}
\caption{Top:  the local pair correlation $g^{(2)}$ as a function of temperature,
for fixed total chemical potential $\mu$ and four different values of the
relative chemical potential $\Omega$.  The non-monotonicity of $g^{(2)}$ in the spinor gas
is to be contrasted to its monotonicity in the Lieb-Liniger case (top, inset).
Bottom:  polarization as a function of temperature, for the same values of
$\Omega$ and $\mu$.  At zero temperature, polarization is total irrespective of
the chemical potential (see also Fig. \ref{fig:T}), illustrating the ferromagnetic-like physics involved in
the spinor gas.}
\label{fig:C_T}
\end{figure}

\begin{figure}
\includegraphics[width=8.5cm]{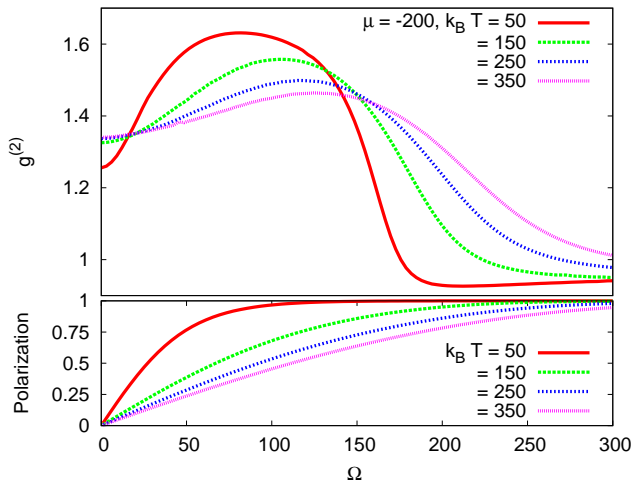}
\caption{Top:  the local pair correlation $g^{(2)}$  as a function of the relative chemical potential
$\Omega$, for fixed total chemical potential $\mu$ and four different values of the
temperature.
Bottom:  polarization as a function of the relative chemical potential, for the same values of
$T$ and $\mu$.}
\label{fig:C_Om}
\end{figure}

When studied as a function of temperature, the 
behaviour of $g^{(2)}$ is markedly different in the two-component gas than in the single-component case.  
For fixed chemical potentials, it can remarkably exhibit a maximum at finite temperature as shown on Fig. \ref{fig:C_T}.  
In the regime of large relative chemical potential, 
the gas is polarized and the correlation is monotonic in temperature; however, one can clearly observe that for $\Omega \rightarrow 0$ 
a peak appears.  For this range of temperatures, $\tau$ is $>1$ but $\gamma$ passes from high values at low $T$ to almost zero for high temperature. 
As a function of the relative chemical potential (Fig. \ref{fig:C_Om} ), $g^{(2)}$ exibits a maximum followed by a local minimum. 
The gas is in the decoherent regime at low relative chemical potential and the saturation value of the correlation for this regime 
increases with $\Omega$. For bigger values $1 \gg \gamma \gg \tau$ and for $\Omega \geq \mu$, the first component is quasi-condensating 
and $g^{(2)} \sim 1$.  

The existence of a maximum and minimum of the density fluctuations is the result of an interesting
competition between interaction energy and entropy.  On the one hand, the bosonic nature
favours ferromagnetic correlations, which in general set the small-temperature thermodynamic properties.  
On the other hand, the reduced entropy associated to the polarized states and the enhanced spatial density 
resulting from quasi-condensation bear a free energy cost which can or cannot be afforded depending on the
value of temperature and of the chemical potentials.  Our work clarifies and quantifies these effects fully
throughout the available parameter space:  depending on the precise values of these three parameters,
we see that the system equilibrates to a state with markedly differing correlations.  As a corollary, the
nonmonotonicity found in $g^{(2)}$ could point to other interesting consequences for experimental realizations of such a system.  
The population densities of a two-component Bose gas in a trap, obtainable by coupling our method to a local density
approximation, would also display interesting correlated behaviour as a function of position.  The ferromagnetic
tendencies of the system would tend to drive phase separation, leading to an observable
correlated enhancement and depletion of the spatial density profiles of the different bosonic species.

The authors would like to thank N.J. van Druten and G. Shlyapnikov for stimulating discussions,
and gratefully acknowledge support from the FOM foundation.

\bibliographystyle{apsrev}
\bibliography{2CBG_Letter}

\end{document}